\RequirePackage[2020-02-02] {latexrelease}
\documentclass[eqsecnum,noshowpacs,showkeys,nofootinbib,aps]{revtex4}

\usepackage{epsfig}
\usepackage{xcolor}
\textcolor{blue}{}
\usepackage{ulem} 

\renewcommand{\theequation}{\arabic{equation}}
\def\be{\begin{equation}}
\def\ee{\end{equation}}
\def\bea{\begin{eqnarray}}
\def\eea{\end{eqnarray}}

\begin{document}

\title{Tidal effects based on GUP-induced effective metric}
\author{Soon-Tae Hong}
\email{galaxy.mass@gmail.com}
\affiliation{Center for Quantum
Spacetime and Department of Physics,
\\  Sogang University, Seoul 04107, Korea}
\author{Yong-Wan Kim}
\email{ywkim65@gmail.com}
 \affiliation{Department of Physics and
Research Institute of Physics and Chemistry, \\ Jeonbuk National
University, Jeonju 54896, Korea}
\author{Young-Jai Park}
\affiliation{Department of Physics, Sogang University, Seoul
04107, Korea }
\date{\today}

\begin{abstract}
In this paper, we study tidal forces in the Schwarzschild black
hole whose metric includes explicitly a generalized uncertainty
principle (GUP) effect. We also investigate interesting features
of the geodesic equations and tidal effects dependent on the GUP
parameter $\alpha$ related to a minimum length. Then, by solving
geodesic deviation equations explicitly with appropriate boundary
conditions, we show that $\alpha$ in the effective metric affects
both the radial and angular components of the geodesic equation,
particularly near the singularities.
\end{abstract}


\maketitle

\section{introduction}
\setcounter{equation}{0}
\renewcommand{\theequation}{\arabic{section}.\arabic{equation}}

Quantum gravity phenomenology predicts the possible existence of a
minimal length on the smallest scale
\cite{Maggiore:1993rv,Kempf:1994su}, which suggests that the
Heisenberg uncertainty principle (HUP) in quantum mechanics should
be modified to use a generalized uncertainty principle (GUP)
\cite{Garay:1994en,Scardigli:1999jh,KalyanaRama:2001xd,
Chang:2001bm,Hossenfelder:2012jw,Tawfik:2014zca}.
In this line of research, various types of GUPs
\cite{Bolen:2004sq,Bambi:2007ty,Park:2007az,Pedram:2011gw,Nozari:2012gd,Chung:2019raj}
have been studied with much interest over the last few decades to
extend the current understanding of general relativity (GR) to the
quantum gravity regime. In particular, using the HUP and GUPs,
quantum effects on classical GR have been explored through the
study of Hawking radiation \cite{Hawking:1974sw}, the thermal
radiation emitted by a black hole. The results heuristically
include the Hawking temperature. Having obtained the Hawking
temperature, one can proceed to study black holes
\cite{Adler:2001vs,Nozari:2005ah,Medved:2004yu,Chen:2014jwq,Carr:2015nqa,Bosso:2023aht}
including their thermodynamics
\cite{Amelino-Camelia:2005zpp,Xiang:2006ei,Myung:2006qr,Hai-Xia:2007,Xiang:2009yq,
Banerjee:2010sd,Gangopadhyay:2013ofa,Gine:2020ves,Lutfuoglu:2021ofc,Su:2022cwd,Yu:2024qzl}
and further to statistical mechanics
\cite{Liu:2001ra,Li:2002xb,Kim:2007nh,Nouicer:2007jg,Kim:2007if,Zhao:2009zzb,Tang:2017wph,Hong:2021xeg}.

However, as is well known, Einstein's GR is the geometric theory
of gravitation and a given spacetime structure is completely
determined by a metric tensor as a solution to Einstein's
equations, which describe the relation between the geometry of a
spacetime and the energy-momentum contained in that spacetime. In
this regard, if one can find a metric tensor including the GUP
effect, it would be more efficient to implement research such as
black hole physics, including both classical and quantum aspects.
Various authors
\cite{Scardigli:2014qka,FaragAli:2015boi,Vagenas:2017vsw,Contreras:2016xib,
Anacleto:2020lel,Anacleto:2021qoe,Jusufi:2022uhk,Chemisana:2023fuk}
have tried to incorporate GUPs into metrics, which we will call a
GUP-induced effective metric. Very recently, Ong
\cite{Ong:2023jkp} has obtained a satisfying GUP-induced effective
metric by emphasizing the importance of using the full GUP
expression instead of just series expansions
\cite{Contreras:2016xib,Anacleto:2020lel,Anacleto:2021qoe}, which
has some subtleties and shortcomings of undesired features.

With the advent of a GUP-induced effective metric, it would be
interesting to study tidal forces and their effects to see what
the modification in the metric gives. In GR, it is well known that
a body in free fall toward the center of another body gets
stretched in the radial direction and compressed in the angular
one \cite{MTW:1973,DInverno:1992,Carroll:2004,Hobson:2006}. These
are due to the tidal effect of gravity that causes two body parts
to be stretched and/or compressed by a difference in the strength
of gravity. These are common in the universe, from our solar
system to stars in binary systems, galaxies, clusters of galaxies,
and even gravitational waves \cite{Goswami:2019fyk}. On a
theoretical side, the tidal effects have been studied in various
spherically symmetric spacetimes, such as Reissner-Nordstr\"om
black hole \cite{Crispino:2016pnv}, Kiselev black hole
\cite{Shahzad:2017vwi}, some regular black holes
\cite{Sharif:2018gaj,Lima:2020wcb}, Schwarzschild black hole in
massive gravity \cite{Hong:2020bdb}, 4D Einstein-Gauss-Bonnet
black hole \cite{Li:2021izh}, Kottler spacetimes
\cite{Vandeev:2021yan}, and many others
\cite{Vandeev:2022gbi,Madan:2022spd,Liu:2022lrg,LimaJunior:2022gko,Abbas:2023gap,
Toshmatov:2023anz}.

In this paper, we study tidal effects produced in the spacetime of
Schwarzschild black hole modified by a GUP-induced effective
metric. In Sec. II, by following Ong's approach, we briefly
recapitulate the method to obtain a GUP-induced effective metric
and study its properties. In Sec. III, we investigate interesting
features of the geodesic equations and in Sec. IV, tidal forces
for a GUP-induced Schwarzschild black hole. Then, we explicitly
find radial and angular solutions of the geodesic deviation
equations for radially falling bodies toward a GUP-induced
Schwarzschild black hole and compare the results with the
Schwarzschild solutions with no GUP effects in Sec. V. Discussion
is drawn in Sec. VI.

\section{GUP-induced effective metric}

In this section, we briefly recapitulate Ong's idea
\cite{Ong:2023jkp} of getting an effective metric based on a GUP
and study the metric's general properties. First of all, we begin
with the most familiar form of a GUP given by
 \be\label{gup}
 \Delta x \Delta p \geq \frac{\hbar}{2}\left(1+\alpha L_{p}^{2}
                        \frac{\Delta p^{2}}{\hbar^2}\right),
 \ee
where $\alpha$ is a dimensionless GUP parameter of order unity and
$L_p$ is the Planck length. It is easy to see that this reduces to
the HUP as $\alpha\rightarrow 0$. The GUP implies the following
inequality
 \be\label{gup-leading-order}
 \frac{\hbar}{\alpha L^2_p}\Delta x\left(1-\sqrt{1-\frac{\alpha L^2_p}{\Delta x^{2}}}\right) \leq \Delta p
 \leq \frac{\hbar}{\alpha L^2_p}\Delta x\left(1+\sqrt{1-\frac{\alpha L^2_p}{\Delta x^{2}}}\right),
 \ee
from which one can find that there exists a minimum bound in the
position uncertainty as
 \be\label{mimL}
 (\Delta x)_{\rm min}=\sqrt{\alpha}L_p.
 \ee

Now, by assuming that photons escape the Schwarzschild black hole
in the radius of $r_H=2M$ and that the spectrum of such escaping
photons is thermal, according to Adler et al. \cite{Adler:2001vs},
one can arrive at the Hawking temperature
 \be\label{gupT}
 T_{\mathrm{GUP}}=\frac{Mc}{\pi \alpha}
 \left(1-\sqrt{1-\frac{\alpha\hbar c}{4GM^{2}}}\right).
 \ee
Here, we have followed the convention of introducing $1/2\pi$
factor to give the Hawking temperature of the Schwarzschild black
hole, $T_{\rm Sch}=1/8\pi M$, in the small $\alpha$ limit. Note
also that $M^2_p=\hbar c/G$.

On the other hand, in order to incorporate the GUP into a metric,
by following Ong's idea \cite{Ong:2023jkp}, one can consider a
modified metric ansatz, without loss of generality, as
 \be\label{f-metric}
 \rm{d}s^{2}=-f(r) \rm{d}t^{2}+f(r)^{-1} \rm{d}r^{2}+r^{2} \rm{d}\Omega^{2}
 \ee
with
 \bea
 f(r)=\left(1-\frac{2GM}{rc^2}\right) g(r),
 \eea
which is the most general form of ansatz while preserving the
areal radius. Then, the Hawking temperature is modified to
 \be\label{gupT-ansatz}
 T=\frac{1}{4\pi}\left.f'(r)\right|_{r=r_H}=\frac{c^2}{8\pi GM}g(r_H),
 \ee
proportional to $g(r_H)$ function. Note that the event horizon
remains intact, {\it i.e.,} at $r_H=2GM/c^2$, $f(r_H)=0$ although
$g(r_H)\neq 0$.
\begin{figure*}[t]
 \centering
 \includegraphics{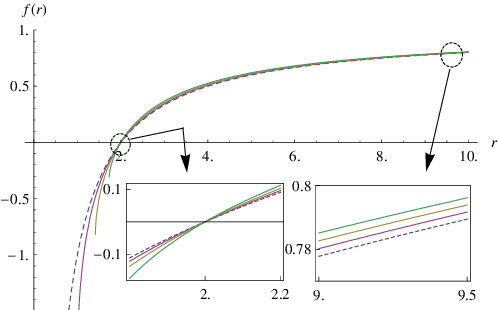}
  \caption{Effective metric: the dashed curve is for $\alpha=0$
  and solid curves are for $\alpha=1.0,~2.0,~3.0$ from down to top.
  Note that we have chosen $M=1$ for figures unless said otherwise. }
 \label{fig1}
\end{figure*}
\begin{figure*}[t]
 \centering
 \includegraphics{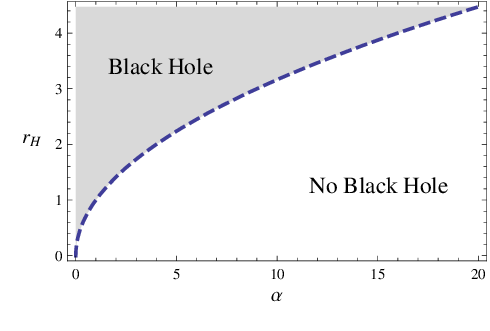}
  \caption{Plot between the event horizon $r_H$ of the black hole and the GUP parameter
  $\alpha$ in the effective metric. }
 \label{fig2}
\end{figure*}
\begin{figure*}[t]
 \centering
 \includegraphics{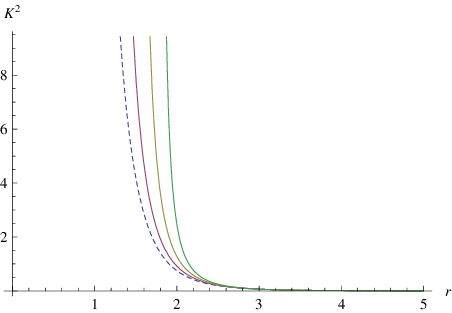}
  \caption{The Kretschmann scalar
  plotted over $r$:
  the dashed curve is for $\alpha=0$
  and solid curves are for $\alpha=1.0,~2.0,~3.0$ from left to right. }
 \label{fig3}
\end{figure*}

By equating the temperature in Eq. (2.7) with the GUP-induced
temperature (2.4), one can have
 \be
 g(r_H)=\frac{2r^2_Hc}{G\alpha}\left(1-\sqrt{1-\frac{\alpha L^2_p}{r^2_H}}\right),
 \ee
where $L^2_p=G\hbar/c^3$. Thus, one can infer the proper form of
$g(r)$ as
 \be
  g(r)=\frac{2r^2c}{G\alpha}\left(1-\sqrt{1-\frac{\alpha L^2_p}{r^2}}\right),
 \ee
which produces an effective metric embodying the effect of the GUP
as
 \be\label{eff-metric0}
 f(r)=\left(1-\frac{2GM}{rc^2}\right)\frac{2r^2c}{G\alpha}\left(1-\sqrt{1-\frac{\alpha L^2_p}{r^2}}\right).
 \ee
For the issue of uniqueness of choosing $g(r)$, we refer Ong's
work \cite{Ong:2023jkp}. Hereafter, for simplicity, we will use
the Planck units of $c=\hbar=G=1$, which also imply that
$L_p=M_p=1$. Thus, the GUP-induced effective metric which we will
study has the form of
  \be\label{eff-metric}
 f(r)=\left(1-\frac{2M}{r}\right)\frac{2r^2}{\alpha}\left(1-\sqrt{1-\frac{\alpha}{r^2}}\right).
 \ee

We have plotted the GUP-induced effective metrics with different
$\alpha$'s in Fig. 1 compared to the original Schwarzschild metric
of $\alpha=0$. This graph shows that the GUP-induced effective
metric functions have the same properties with the Schwarzschild
case both at the asymptotic infinity and at the event horizon. On
the other hand, in between the regions of $r_H< r<\infty$, the
curves of the GUP-induced effective metrics are slightly higher
than the original Schwarzschild metric case, and when $r<r_H$,
they are lower, as shown in the inserted boxes in Fig. 1. In
particular, since the metric in Eq. (2.11) is physically
meaningful when $r\ge\sqrt{\alpha}$ due to the square root, the
GUP parameter $\alpha$ makes black hole solutions possible when
$\sqrt{\alpha}<r_H$, which was drawn in Fig. 2. Therefore, in this
paper, we have concentrated on the GUP parameter $\alpha$ in the
range of $0\le\alpha\le r^2_H$ to study GUP modification effects
on the Schwarzschild black hole. Finally, in the small $\alpha$
limit, the GUP-induced effective metric is reduced to
 \be
 f(r)=\left(1-\frac{2M}{r}\right)\left(1+\frac{\alpha}{4r^2}+\frac{\alpha^2}{8r^4}\right)
 \ee
up to $\alpha^2$-orders. From the second parenthesis multiplied by
the original Schwarzschild metric, one may infer how much the
GUP-induced effective metric differs qualitatively from the
original Schwarzschild metric in the whole range of $r$.

On the other hand, the Kretschmann scalar for the effective metric
is given by
 \be
 K^2 \equiv K_{\mu\nu\rho\sigma} K^{\mu\nu\rho\sigma}
            =\frac{1}{r^3(r^2-\alpha)^2\alpha^2}\left(\frac{k_1}{r(r^2-\alpha)}
            -16k_2\sqrt{1-\frac{\alpha}{r^2}}\right),
 \ee
where
 \bea
 k_1 &=& 192r^{10}-384Mr^9+32(8M^2-19\alpha)r^8+1216M\alpha r^7
         -8(96M^2-85\alpha)\alpha r^6-1360M\alpha^2 r^5\nonumber\\
      &&+4(212M^2-79\alpha)\alpha^2 r^4+608M\alpha^3 r^3-12(32M^2-5\alpha)\alpha^3 r^2 -96M \alpha^4 r +4(16M^2-\alpha)\alpha^4,\nonumber\\
 k_2 &=&12r^7-24Mr^6+4(4M^2-5\alpha)r^5+40M\alpha r^4-8(3M^2-\alpha)\alpha r^3-16M\alpha^2 r^2\nonumber\\
      && +(8M^2-\alpha)\alpha^2 r+2M\alpha^3.
 \eea
One can find that there is no curvature singularity anywhere
except $r=0$ and $r=\sqrt{\alpha}$. Note that in the limit of
$\alpha\rightarrow 0$, the Kretschmann scalar recovers the
Schwarzschild case as
 \be
 K^2 = \frac{48M^2}{r^6}.
 \ee
In Fig. 3, we have drawn the Kretschmann scalar for the GUP
modified effective metric showing that there is no curvature
singularity except $r=0$ and $r=\sqrt{\alpha}$.

\section{Geodesic in Schwarzschild black hole with effective metric}

Now, from the GUP-induced effective metric (2.5) with (2.11), one
can calculate the geodesic equations of
 \be\label{gdeq}
 \frac{\rm{d}^2x^\mu}{\rm{d}\tau^2}+\Gamma^\mu_{\nu\rho}\frac{\rm{d}x^\nu}{\rm{d}\tau}\frac{\rm{d}x^\rho}{\rm{d}\tau}=0,
 \ee
where $x^\mu=(t,r,\theta,\phi)$. With the non-vanishing components
of the Christoffel symbols
 \bea\label{csymbols}
 \Gamma^0_{01}&=&-\Gamma^1_{11}=\frac{f'(r)}{2f(r)},
   ~~~\Gamma^1_{00}=\frac{1}{2}f'(r)f(r),
   ~~~~~\Gamma^1_{22}=-r f(r),
   ~~~\Gamma^1_{33}=-r f(r)\sin^2\theta, \nonumber\\
   \Gamma^2_{12}&=&\Gamma^3_{13}=\frac{1}{r},
   ~~~~~~~~~~~\Gamma^2_{33}=-\sin\theta\cos\theta,
   ~~~~\Gamma^3_{23}=\cot\theta,
 \eea
one can explicitly obtain the geodesic equations as
 \bea
 &&\frac{\rm{d}v^0}{\rm{d}\tau}-\frac{\left(1-\frac{2M}{r}-\sqrt{1-\frac{\alpha}{r^2}}\right)}{r\left(1-\frac{2M}{r}\right)\sqrt{1-\frac{\alpha}{r^2}}}v^0v^1=0, \\
 &&\frac{\rm{d}v^1}{\rm{d}\tau}+\frac{2r\left(1-\frac{2M}{r}\right)\left(1-\sqrt{1-\frac{\alpha}{r^2}}\right)
                       \left(\alpha-2r^2\left(1-\frac{M}{r}\right)\left(1-\sqrt{1-\frac{\alpha}{r^2}}\right)\right)}
                      {\alpha^2\sqrt{1-\frac{\alpha}{r^2}}}(v^0)^2
      +\frac{\left(1-\frac{2M}{r}-\sqrt{1-\frac{\alpha}{r^2}}\right)}{2r\left(1-\frac{2M}{r}\right)\sqrt{1-\frac{\alpha}{r^2}}}(v^1)^2
      \nonumber\\
      &&-r\left(1-\frac{2M}{r}\right)\frac{2r^2}{\alpha}\left(1-\sqrt{1-\frac{\alpha}{r^2}}\right)[(v^2)^2+\sin^2\theta(v^3)^2]=0,     \\
 &&\frac{\rm{d}v^2}{\rm{d}\tau}+\frac{2}{r}v^1v^2-\sin\theta\cos\theta(v^3)^2=0, \\
 &&\frac{\rm{d}v^3}{\rm{d}\tau}+\frac{2}{r}v^1v^3+2\cot\theta v^2 v^3=0,
 \eea
where we denote the four-velocity vector as
$v^\mu=\rm{d}x^\mu/\rm{d}\tau$. For simplicity, one can consider
the geodesics on the equatorial plane $\theta=\pi/2$ and thus
$v^2=\dot\theta=0$ for all $\tau$, without loss of generality.
Then, the geodesic equations are simplified to
 \bea
 &&\frac{\rm{d}v^0}{\rm{d}\tau}-\frac{\left(1-\frac{2M}{r}-\sqrt{1-\frac{\alpha}{r^2}}\right)}{r\left(1-\frac{2M}{r}\right)\sqrt{1-\frac{\alpha}{r^2}}}v^0v^1=0, \label{ge0a}\\
 &&\frac{\rm{d}v^1}{\rm{d}\tau}+\frac{2r\left(1-\frac{2M}{r}\right)\left(1-\sqrt{1-\frac{\alpha}{r^2}}\right)
                       \left(\alpha-2r^2\left(1-\frac{M}{r}\right)\left(1-\sqrt{1-\frac{\alpha}{r^2}}\right)\right)}
                      {\alpha^2\sqrt{1-\frac{\alpha}{r^2}}}(v^0)^2
      +\frac{\left(1-\frac{2M}{r}-\sqrt{1-\frac{\alpha}{r^2}}\right)}{2r\left(1-\frac{2M}{r}\right)\sqrt{1-\frac{\alpha}{r^2}}}(v^1)^2
      \nonumber\\
      &&-r\left(1-\frac{2M}{r}\right)\frac{2r^2}{\alpha}\left(1-\sqrt{1-\frac{\alpha}{r^2}}\right)(v^3)^2=0,     \\
 &&\frac{\rm{d}v^3}{\rm{d}\tau}+\frac{2}{r}v^1v^3=0.\label{ge3a}
 \eea
It is now easy to find solutions for Eqs. (3.7) and (3.9) by
direct integrations as
 \bea
 v^0&=&\frac{c_1}{\left(1-\frac{2M}{r}\right)\frac{2r^2}{\alpha}
        \left(1-\sqrt{1-\frac{\alpha}{r^2}}\right)},\label{v0aF} \\
 v^3&=&\frac{c_2}{r^2}, \label{v3aF}
 \eea
respectively, where $c_1$ and $c_2$ are integration constants. Two
conserved quantities defined by
 \bea
 E &=& -g_{\mu\nu}\xi^\mu v^\nu,\\
 L &=& g_{\mu\nu}\psi^\mu v^\nu
 \eea
can be used to identify the integration constants $c_1$, $c_2$
with $E$, $L$, respectively. Here, $\xi^\mu=(1,0,0,0)$ and
$\psi^\mu=(0,0,0,1)$ are the Killing vectors. Finally, by letting
$\rm{d}s^2=-k\rm{d}\tau^2$ and using Eqs. (3.10) and (3.11), one
can obtain
 \be
 v^1=\frac{\rm{d}r}{\rm{d}\tau}=\pm\left[E^2-\left(k+\frac{L^2}{r^2}\right)
 \left(1-\frac{2M}{r}\right)\frac{2r^2}{\alpha}\left(1-\sqrt{1-\frac{\alpha}{r^2}}\right)\right]^{1/2},
 \ee
where $+/-$ sign is for outward/inward motion. Also, timelike
(nulllike) geodesic is for $k=1~(0)$.

\section{Tidal force in Schwarzschild black hole with effective metric}
\setcounter{equation}{0}
\renewcommand{\theequation}{\arabic{section}.\arabic{equation}}

Now, let us investigate the tidal force acting in the
Schwarzschild black hole modified by the effective metric. First
of all, let us consider the geodesic deviation equation
\cite{MTW:1973,DInverno:1992,Carroll:2004,Hobson:2006}
 \be\label{gd}
 \frac{\rm{D}^2\eta^\mu}{\rm{D}\tau^2}+R^\mu_{\nu\rho\sigma}v^\nu
 \eta^\rho v^\sigma = 0,
 \ee
where $R^\mu_{\nu\rho\sigma}$ is the Riemann curvature and $v^\mu$
is the unit tangent vector to the geodesic line. The geodesic
deviation equation describes the behavior of a one-parameter
family of neighboring geodesics through relative separation
four-vectors $\eta^\mu$, the infinitesimal displacement between
two nearby geodesics.

In order to study the behavior of the separation vectors in
detail, we consider the timelike geodesic equation with $L=0$ for
simplicity. We also introduce the tetrad basis describing a freely
falling frame given by
 \bea
 e^\mu_{\hat 0}&=&\left(\frac{E}{f(r)},-\sqrt{E^2-f(r)},0,0\right),\nonumber\\
 e^\mu_{\hat 1}&=&\left(-\frac{\sqrt{E^2-f(r)}}{f(r)},E,0,0\right),\nonumber\\
 e^\mu_{\hat 2}&=&\left(0,0,\frac{1}{r},0\right),\nonumber\\
 e^\mu_{\hat 3}&=&\left(0,0,0,\frac{1}{r\sin\theta}\right),
 \eea
satisfying the orthonormality relation of $e^\mu_{\hat\alpha}
e_{\mu\hat\beta}=\eta_{\hat\alpha\hat\beta}$ with
$\eta_{\hat\alpha\hat\beta}={\rm diag}(-1,1,1,1)$. The separation
vectors can also be expanded as
$\eta^\mu=e^\mu_{\hat\alpha}\eta^{\hat\alpha}$ with a fixed
temporal component of $\eta^{\hat 0}=0$
\cite{DInverno:1992,Hobson:2006}.

In the tetrad basis, the Riemann tensor can be written as
 \be
 R^{\hat\alpha}_{\hat\beta\hat\gamma\hat\delta}
 =e^{\hat\alpha}_\mu e^\nu_{\hat\beta} e^\rho_{\hat\gamma}
 e^\sigma_{\hat\delta} R^\mu_{\nu\rho\sigma},
 \ee
so one can obtain the non-vanishing independent components of the
Riemann tensor in the Schwarzschild black hole modified by the
effective metric as
 \bea
 R^{\hat 0}_{\hat 1 \hat 0 \hat 1}&=&-\frac{f''(r)}{2},
    ~~R^{\hat 0}_{\hat 2 \hat 0 \hat 2}=
      R^{\hat 0}_{\hat 3 \hat 0 \hat 3}=
      R^{\hat 1}_{\hat 2 \hat 1 \hat 2}=
      R^{\hat 1}_{\hat 3 \hat 1 \hat 3}=-\frac{f'(r)}{2r},
    ~~R^{\hat 2}_{\hat 3 \hat 2 \hat 3}=\frac{1-f(r)}{r^2}.
 \eea
Then, one can obtain the desired tidal forces in the radially
freely falling frame as
 \bea
 \frac{\rm{d}^2\eta^{\hat 1}}{\rm{d}\tau^2} &=& -\frac{f''(r)}{2}\eta^{\hat 1}
                 =\frac{(2M-3r)\alpha+2r^3\left(1-\left(1-\frac{\alpha}{r^2}\right)^{3/2}\right)}{r^3\alpha\left(1-\frac{\alpha}{r^2}\right)^{3/2}}\eta^{\hat 1},\label{gdem1}\\
 \frac{\rm{d}^2\eta^{\hat i}}{\rm{d}\tau^2} &=& -\frac{f'(r)}{2r}\eta^{\hat i}
                = -\frac{\alpha+2(M-r)r\left(1-\sqrt{1-\frac{\alpha}{r^2}}\right)}{r^2\alpha\sqrt{1-\frac{\alpha}{r^2}}}\eta^{\hat i},\label{gdem2}
 \eea
where $i=2,~3$. As $\alpha\rightarrow 0$, they are reduced to
 \bea
 \frac{\rm{d}^2\eta^{\hat 1}}{\rm{d}\tau^2} &=& \frac{2M}{r^3}\eta^{\hat 1},\\
 \frac{\rm{d}^2\eta^{\hat i}}{\rm{d}\tau^2} &=& -\frac{M}{r^3}\eta^{\hat i},
 \eea
which are the radial and angular tidal forces of the Schwarzschild
black hole. As a result, we have newly obtained the tidal effect
dependent on the GUP parameter $\alpha$ that is embodied in the
effective metric. In Fig. 4, we have plotted the radial and
angular tidal forces by comparing them with the original
Schwarzschild case.
\begin{figure*}[t]
 \centering
 \includegraphics{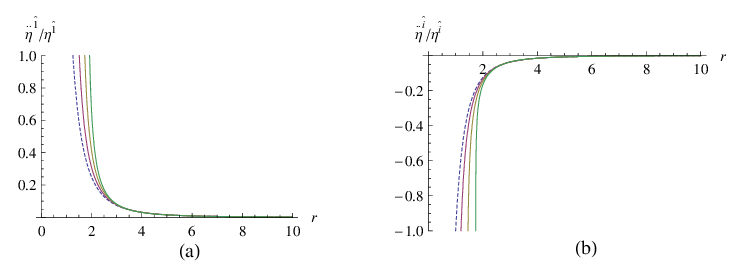}
  \caption{(a) Radial and (b) angular tidal forces with different GUP parameters
  $\alpha=1.0,~2.0,~3.0$ from the left.
   The dashed curves are for $\alpha=0$, the original Schwarzschild case. }
 \label{fig4}
\end{figure*}
As seen in Fig. 4(a), the radial tidal forces are always positive,
and as the GUP parameters $\alpha$ increase, the curves move to
the right while the curve's shapes remain the same. It seems
appropriate to comment that the radial tidal forces go to
infinities as they approach their singularities, so the radial
stretchings get infinities. The similar interpretation can be
applied to the angular tidal forces. However, note that they go to
negative infinities as $r$ approaches the singularity, as seen in
Fig. 4(b). Thus, there are infinite compressions in the angular
direction.

\section{Geodesic deviation equations of Schwarzschild black hole with effective metric}
\setcounter{equation}{0}
\renewcommand{\theequation}{\arabic{section}.\arabic{equation}}

For the geodesic deviation equations (4.5) and (4.6) of the
Schwarzschild black hole modified by the effective metric, the
tidal forces can be rewritten in terms of $r$-derivative as
 \bea\label{diffm1}
 [E^2-f(r)]\frac{\rm{d}^2\eta^{\hat 1}}{\rm{d}r^2}
  - \frac{f'(r)}{2}\frac{\rm{d}\eta^{\hat 1}}{\rm{d}r}+\frac{f''(r)}{2}\eta^{\hat
  1}&=&0,\\
 \label{diffm2}
 [E^2-f(r)]
 \frac{\rm{d}^2\eta^{\hat i}}{\rm{d}r^2}
  - \frac{f'(r)}{2}\frac{\rm{d}\eta^{\hat i}}{\rm{d}r}+\frac{f'(r)}{2r}\eta^{\hat i}&=&0.
 \eea
The solution of the radial component (5.1) is known to have the
following general form of
 \be\label{rintegral}
 \eta^{\hat 1}(r)= c_1\sqrt{E^2-f(r)}
       +c_2\sqrt{E^2-f(r)}\int\frac{\rm{d}r}{[E^2-f(r)]^{3/2}},
 \ee
and for the angular component (5.2), it is
 \be\label{aintegral}
 \eta^{\hat i}(r)=
    r\left(c_3+c_4\int\frac{\rm{d}r}{r^2\sqrt{E^2-f(r)}}\right),
 \ee
where $c_i~(i=1,2,3,4) $ are constants of integration
\cite{Crispino:2016pnv,Shahzad:2017vwi,Sharif:2018gaj,Lima:2020wcb,Hong:2020bdb,Li:2021izh,Vandeev:2021yan},
which will be determined by the boundary conditions.

Now, let us solve the geodesic deviation equations (5.1) and (5.2)
by series expansions of the integrands to the power of $\alpha$.
Firstly, we consider a body released from rest at $r=b$ so that we
have $E=\sqrt{f(b)}$. Then, the term $E^2-f(r)$ can be expressed
as
 \be
 E^2-f(r)=2M\left(\frac{1}{r}-\frac{1}{b}\right)Q(r),
 \ee
where
 \bea
 Q(r)&=&1+\sum^{\infty}_{n=1}\alpha^n\frac{(2n-1)!!}{(n+1)!2^n}
               \left(-\frac{P_n(r)}{2M}+P_{n+1}(r)\right),\nonumber\\
       &\equiv& 1+\sum^{\infty}_{n=1}\alpha^nQ^{(n)}(r),   \nonumber\\
 P_n(r)&=& \sum^{n}_{m=0}\frac{1}{r^{n-m}b^m}.
 \eea
Here, $Q^{(n)}(r)$ is the function of $r$ of order $\alpha^n$.
Then, without loss of generality, the solution of the radial
component (5.1) can be obtained up to $\alpha^2$ order as
 \bea
 \eta^{\hat 1}(r)&=&c_1\sqrt{\frac{2M}{b}}\frac{\sqrt{br-r^2}}{r}\sqrt{1+\alpha Q^{(1)}(r)+\alpha^2Q^{(2)}(r)}  \nonumber\\
              &+& c_2\frac{b}{2M}\sqrt{1+\alpha Q^{(1)}(r)+\alpha^2 Q^{(2)}(r)}
                \left[2bd_1(r)+\frac{3b}{2r}d_2(r)\sqrt{br-r^2}\cos^{-1}\left(\frac{2r}{b}-1\right)\right],\\
 \eta^{\hat i}(r)&=& c_3 r
               -c_4\sqrt{\frac{2b}{M}}\frac{\sqrt{br-r^2}}{b}d_3(r),
 \eea
where
 \bea
 d_1(r)&=& \frac{3}{2}-\frac{r}{2b}+\frac{3\alpha}{16b}\left(\frac{5}{2M}-\frac{7}{b}+\frac{r}{b^2}-\frac{r}{2bM}\right)\nonumber\\
       &+& \frac{3\alpha^2}{64b^2r}\left(\frac{1}{b}+\frac{1}{2M}+\frac{1}{2r}+\frac{r}{4b^2}+\frac{45r}{16M^2}
            -\frac{31r}{4bM}+\frac{3r^2}{4b^3}-\frac{5r^2}{16bM^2}+\frac{r^2}{4b^2M}\right), \nonumber\\
 d_2(r)&=& 1+\frac{5\alpha}{8b}\left(\frac{1}{2M}-\frac{1}{b}\right)+\frac{5\alpha^2}{128b^2}\left(\frac{7}{4M^2}-\frac{3}{bM}-\frac{1}{b^2}\right),\nonumber\\
 d_3(r)&=& 1+\frac{\alpha}{8}\left(\frac{5}{6Mb}-\frac{11}{5b^2}+\frac{1}{6Mr}-\frac{3}{5br}-\frac{1}{5r^2}\right)
           +\frac{\alpha^2}{16}\left(\frac{43}{160b^2M^2}-\frac{33}{280b^3M}-\frac{551}{504b^4}+\frac{7}{80bM^2r}\right. \nonumber\\
       &-& \left.\frac{17}{140b^2Mr}-\frac{59}{252b^3r}+\frac{3}{160M^2r^2}-\frac{1}{35bMr^2}-\frac{19}{168b^2r^2}
           +\frac{1}{56Mr^3}-\frac{29}{252br^3}-\frac{5}{72r^4}\right).
 \eea
Note that $d_2(r)$ has no $r$-dependence. The integration
constants $c_i$ can be determined by adopting appropriate boundary
conditions that the infinitesimal displacement and initial
velocity between two nearby particles are given by $\eta^k(b)$ and
$\rm{d}\eta^k(r)/\rm{d}\tau|_{r=b}\equiv
\rm{d}\eta^k(b)/\rm{d}\tau$ ($k=1,~2,~3$) at $r=b$. Explicitly, we
have
 \bea
 c_1&=&\frac{b^2}{M}\left(1+\alpha Q^{(1)}(b)+\alpha^2 Q^{(2)}(b)\right)^{-1}\frac{\rm{d}\eta^{\hat 1}(b)}{\rm{d}\tau},\nonumber\\
 c_2&=&\frac{M}{b^2}\left(1+\alpha Q^{(1)}(b)+\alpha^2 Q^{(2)}(b)\right)^{-1/2}d^{-1}_1(b)\eta^{\hat 1}(b),\nonumber\\
 c_3&=&\frac{1}{b}\eta^{\hat i}(b),\nonumber\\
 c_4&=&-b\left(1+\alpha Q^{(1)}(b)+\alpha^2 Q^{(2)}(b)\right)^{-1/2}d^{-1}_3(b)\frac{\rm{d}\eta^{\hat i}(b)}{\rm{d}\tau}.
 \eea
Thus, the solutions are finally written as
 \bea
 \eta^{\hat 1}(r)&=&b\sqrt{\frac{2b}{M}}\frac{\rm{d}\eta^{\hat 1}(b)}{\rm{d}\tau}
                     \frac{\sqrt{br-r^2}}{r}
                     \frac{\left(1+\alpha Q^{(1)}(r)+\alpha^2 Q^{(2)}(r)\right)^{1/2}}{\left(1+\alpha Q^{(1)}(b)+\alpha^2 Q^{(2)}(b)\right)}  \nonumber\\
              &+& \eta^{\hat 1}(b)d^{-1}_1(b)
              \left(\frac{1+\alpha Q^{(1)}(r)+\alpha^2 Q^{(2)}(r)}{1+\alpha Q^{(1)}(b)+\alpha^2 Q^{(2)}(b)}\right)^{1/2}
                \left[d_1(r)+\frac{3}{4r}d_2(r)\sqrt{br-r^2}\cos^{-1}\left(\frac{2r}{b}-1\right)\right],\\
 \eta^{\hat i}(r)&=& \frac{1}{b}\eta^{\hat i}(b) r
               +\sqrt{\frac{2b}{M}}\frac{\rm{d}\eta^{\hat i}(b)}{\rm{d}\tau}d^{-1}_2(b)\left(\frac{br-r^2}{1+\alpha Q^{(1)}(b)+\alpha^2 Q^{(2)}(b)}\right)^{1/2}d_3(r),
 \eea
\begin{figure*}[t]
 \centering
 \includegraphics{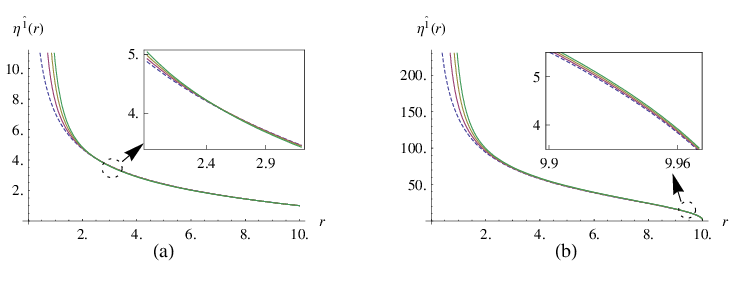}
  \caption{Radial components of geodesic deviation in the Schwarzschild black hole
  modified by an effective metric, (a) with $\frac{\rm{d}\eta^{\hat{1}}(b)}{\rm{d}\tau}=0$, and
  (b) with $\frac{\rm{d}\eta^{\hat{1}}(b)}{\rm{d}\tau}=1~(\neq 0)$, the dashed curves are for $\alpha=0$, and
  the solid curves for $\alpha=1.0,~2.0,~3.0$ from the left.}
 \label{fig5}
\end{figure*}
\begin{figure*}[t]
 \centering
 \includegraphics{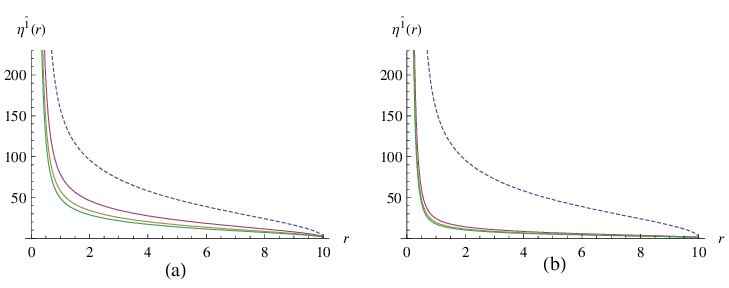}
  \caption{Radial components of geodesic deviation by varying the
  mass $M$ in the Schwarzschild black hole modified by an effective metric
  with $\frac{\rm{d}\eta^{\hat{1}}(b)}{\rm{d}\tau}=1$.
  (a) the solid curves for $M=5,~10,~15$, (b) the solid curves for $M=100,~200,~300$, downwards,
  with the dashed curve for $M=1$. }
 \label{fig6}
\end{figure*}
\begin{figure*}[t]
 \centering
 \includegraphics{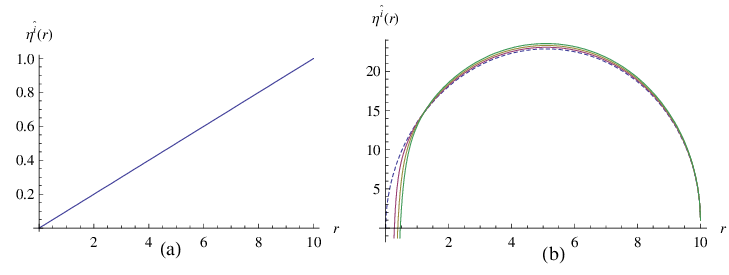}
  \caption{Angular components of geodesic deviation in the Schwarzschild black hole
  modified by the $\alpha$ dependent effective metric, (a) with $\alpha=0$ and $\frac{\rm{d}\eta^{\hat{1}}(b)}{\rm{d}\tau}=0$,
  (b) with $\alpha=0$ and $\frac{\rm{d}\eta^{\hat{1}}(b)}{\rm{d}\tau}=1$ for the dashed curve, and $\alpha=1.0,~2.0,~3.0$
  (from the left near $r=0.5$) and $\frac{\rm{d}\eta^{\hat{1}}(b)}{\rm{d}\tau}=1$ for the solid curves.}
 \label{fig7}
\end{figure*}
\begin{figure*}[t]
 \centering
 \includegraphics{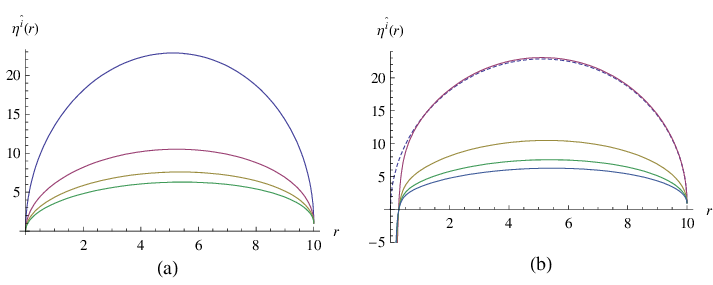}
  \caption{(a) Angular components of geodesic deviation by varying the mass $M$ in the effective metric
  for $M=1,~5,~10,~15$ (from top to down) with $\alpha=0$ in the original Schwarzschild black hole,
  (b) Angular components of geodesic deviation by varying the mass in the effective metric
  for $M=1,~5,~10,~15$ with $\alpha=1$ in the Schwarzschild black hole modified by an effective metric.
  Here, the dashed curve is for $\alpha=0$ case, drawn for comparison.}
 \label{fig8}
\end{figure*}
\begin{figure*}[t]
 \centering
 \includegraphics{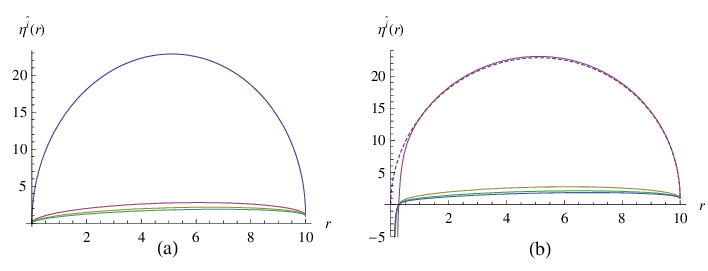}
  \caption{(a) Angular components of geodesic deviation by varying the mass $M$ in the effective metric
  for $M=1,~100,~200,~300$ (from top to down) with $\alpha=0$ in the original Schwarzschild black hole,
  (b) Angular components of geodesic deviation by varying the mass in the effective metric
  for $M=1,~100,~200,~300$ with $\alpha=1$ in the Schwarzschild black hole modified by an effective metric.
  Here, the dashed curve is for $\alpha=0$ case, as before, drawn for comparison.}
 \label{fig9}
\end{figure*}
In Fig. 5, we have drawn the radial components of the geodesic
deviation by comparing the case of not having GUP parameter to
that of having a GUP parameter for both
$\frac{\rm{d}\eta^{\hat{1}}(b)}{\rm{d}\tau}=0$ and
$\frac{\rm{d}\eta^{\hat{1}}(b)}{\rm{d}\tau}\neq 0$ initial
velocities. From the figure, one can see that the radial
separation vectors of a falling object in the GUP-modified
Schwarzschild black hole start with the same initial values as the
original Schwarzschild black hole, and then it gets stretched
steeper than the original Schwarzschild case as it approaches the
singularities. This results in much greater separation effects.
Note that comparing Fig. 5(a) with Fig. 5(b), the latter shows a
much steeper slope in the range of $0<r<10$ than the former. In
Fig. 6, we have also drawn the radial components of the geodesic
deviation by varying the mass parameter $M$ in the Schwarzschild
black hole modified by the effective metric. Here, one can find
that the bigger the black hole mass is, the greater the geodesic
separation is near the singularity. Moreover, for comparison, we
have drawn the radial components of geodesic deviation for the
Planck mass comparable scales $M$ in Fig. 6(a) and for some larger
$M$ in Fig. 6(b), which shows that the latter sharply increases
compared to the former as it approaches the singularity.

On the other hand, in Fig. 7, we have drawn the angular components
of the geodesic deviation in the Schwarzschild black hole modified
by the $\alpha$ dependent effective metric. Fig. 7(a) corresponds
to $\alpha=0$ with zero initial velocity,
$\frac{\rm{d}\eta^{\hat{1}}(b)}{\rm{d}\tau}=0$, the original
Schwarzschild case. When the initial velocity of the original
Schwarzschild black hole is non-zero as
$\frac{\rm{d}\eta^{\hat{1}}(b)}{\rm{d}\tau}\neq 0$, the angular
separation is drawn by the dashed curve in Fig. 7(b). Now, the
angular components of the geodesic deviation in the Schwarzschild
black hole modified by an effective metric by varying the GUP
parameter $\alpha$ are drawn by the solid curves in Fig. 7(b).
Note that as the GUP parameters $\alpha$ increase, the angular
separation appears to be larger in all ranges except near the
singularity, where the angular separation is inverted to be small.
Moreover, in Fig. 8, we have also drawn the angular components of
the geodesic deviation by varying the mass. The original
Schwarzschild case is drawn in Fig. 8(a). The GUP-modified
Schwarzschild black hole case is drawn in Fig. 8(b), which shows
that it resembles the original Schwarzschild case but differs near
the singularities. Moreover, for purposes of comparison, for some
large $M$ cases, we have drawn the angular components of geodesic
deviation in Fig. 9, which shows the angular separations are not
much changed.

It seems appropriate to comment that as it approaches the
singularity, the radial and angular components of the geodesic
deviation of the GUP-modified Schwarzschild black hole are
stretched steeper and larger than the ones of the original
Schwarzschild black hole, respectively. It looks peculiar since we
usually expect quantum gravity effects such as a GUP to make
physical quantities more finite, not worse. However, there is also
an example of applying GUP to white dwarfs, which leads to a
peculiar property. A GUP can remove the Chandrasekhar limit
\cite{Rashidi:2015rro,Ong:2018zqn,Mathew:2020wnx}, and thus white
dwarfs can be seemingly arbitrarily large. In this respect, a GUP
correction can lead to more divergent behaviors for some physical
quantities, which goes against our usual expectations and should
be further studied.

\section{Discussion}

In this paper, we have studied tidal effects in a GUP effect
embodied Schwarzschild black hole. We have first recapitulated the
GUP-induced effective metric by following Ong's approach and have
studied its properties in detail. As a result, we have shown that
the $\alpha$ dependent effective metric is only valid in the $0\le
\alpha \le r^2_H$ range. When $\alpha>r^2_H$, it does not even
form an event horizon. Moreover, we have also obtained the
Kretschmann scalar for the effective metric, showing no curvature
singularity except $r=0$ and $r=\sqrt{\alpha}$.

Then, we have investigated interesting features of the geodesic
equations and tidal forces dependent on the GUP parameter
$\alpha$. By comparing the radial and angular tidal forces with
the original Schwarzschild case, we have shown that as the GUP
parameters $\alpha$ increase, the radial and angular tidal forces
go to infinities faster than the original Schwarzschild black
hole's ones.

Furthermore, by considering a free fall of a body released from
rest at $r=b$, we have derived the geodesic deviation equations
and explicitly and analytically solved them. As a result, we have
shown that the radial components of the geodesic deviation get
stretched steeper than the original Schwarzschild case as
approaching the singularities, resulting in much greater
separation effects. Also, by varying the mass parameter $M$, we
have found that the bigger the black hole mass is, the greater the
geodesic separation is near the singularity. On the other hand, as
the GUP parameters $\alpha$ increase, the curves of the angular
components of the geodesic deviation are higher over almost all
ranges $r<b$, and then they become lower near the singularity than
the original Schwarzschild case. As a result, we have found that
the $\alpha$ dependent effective metric affects both the radial
and angular components, particularly near the singularities,
compared with the original Schwarzschild black hole case.

Finally, it seems appropriate to comment that our coordinate
systems still contains a singularity at $r=\sqrt{\alpha}$,
implying that the geodesics is still incomplete inside the black
hole. In this respect, it would be interesting to construct and
analyze the tidal effect, related to the works
\cite{Dvali:2010bf,Isi:2013cxa,Li:2016yfd,Alencar:2023wyf}, on
GUP-induced geodesically compete non-singular black holes through
further investigation.

\acknowledgments{S. T. H. was supported by Basic Science Research
Program through the National Research Foundation of Korea funded
by the Ministry of Education, NRF-2019R1I1A1A01058449. Y. W. K.
was supported by the National Research Foundation of Korea (NRF)
grant funded by the Korea government (MSIT) (No.
2020R1H1A2102242). }



\begin{thebibliography}{99}


\bibitem{Maggiore:1993rv}
M.~Maggiore, A Generalized uncertainty principle in quantum
gravity, Phys. Lett. B \textbf{304}, 65-69 (1993)
[arXiv:hep-th/9301067 [hep-th]].


\bibitem{Kempf:1994su}
A.~Kempf, G.~Mangano and R.~B.~Mann,Hilbert space representation
of the minimal length uncertainty relation, Phys. Rev. D
\textbf{52}, 1108-1118 (1995)
[arXiv:hep-th/9412167 [hep-th]].

\bibitem{Garay:1994en}
L.~J.~Garay, Quantum gravity and minimum length, Int. J. Mod.
Phys. A \textbf{10}, 145 (1995)
[arXiv:gr-qc/9403008 [gr-qc]].

\bibitem{Scardigli:1999jh}
F.~Scardigli, Generalized uncertainty principle in quantum gravity
from micro-black hole Gedanken experiment, Phys. Lett. B
\textbf{452}, 39 (1999)
[arXiv:hep-th/9904025 [hep-th]].

\bibitem{KalyanaRama:2001xd}
S.~Kalyana Rama, Some consequences of the generalized uncertainty
principle: Statistical mechanical, cosmological, and varying speed
of light, Phys. Lett. B \textbf{519}, 103 (2001)
[arXiv:hep-th/0107255 [hep-th]].

\bibitem{Chang:2001bm}
L.~N.~Chang, D.~Minic, N.~Okamura and T.~Takeuchi, The Effect of
the minimal length uncertainty relation on the density of states
and the cosmological constant problem, Phys. Rev. D \textbf{65},
125028 (2002)
[arXiv:hep-th/0201017 [hep-th]].

\bibitem{Hossenfelder:2012jw}
S.~Hossenfelder, Minimal length scale scenarios for quantum
gravity, Living Rev. Rel. \textbf{16}, 2 (2013)
[arXiv:1203.6191 [gr-qc]].

\bibitem{Tawfik:2014zca}
A.~N.~Tawfik and A.~M.~Diab, Generalized uncertainty principle:
approaches and applications, Int. J. Mod. Phys. D \textbf{23},
1430025 (2014)
[arXiv:1410.0206 [gr-qc]].


\bibitem{Bolen:2004sq}
B.~Bolen and M.~Cavaglia, (Anti-)de Sitter black hole
thermodynamics and the generalized uncertainty principle, Gen.
Rel. Grav. \textbf{37}, 1255 (2005)
[arXiv:gr-qc/0411086 [gr-qc]].

\bibitem{Bambi:2007ty}
C.~Bambi and F.~R.~Urban, Natural extension of the generalised
uncertainty principle, Class. Quant. Grav. \textbf{25}, 095006
(2008)
[arXiv:0709.1965 [gr-qc]].

\bibitem{Park:2007az}
M.~I.~Park, The generalized uncertainty principle in (A)dS space
and the modification of Hawking temperature from the minimal
length, Phys. Lett. B \textbf{659}, 698 (2008)
[arXiv:0709.2307 [hep-th]].

\bibitem{Pedram:2011gw}
P.~Pedram, A higher order GUP with minimal length uncertainty and
maximal momentum, Phys. Lett. B \textbf{714}, 317 (2012)
[arXiv:1110.2999 [hep-th]].

\bibitem{Nozari:2012gd}
K.~Nozari and A.~Etemadi, Minimal length, maximal momentum and
Hilbert space representation of quantum mechanics, Phys. Rev. D
\textbf{85}, 104029 (2012)
[arXiv:1205.0158 [hep-th]].

\bibitem{Chung:2019raj}
W.~S.~Chung and H.~Hassanabadi, A new higher order GUP: one
dimensional quantum system, Eur. Phys. J. C \textbf{79}, 213
(2019).





\bibitem{Hawking:1974sw}
S.~W.~Hawking, Particle creation by black holes, Commun. Math.
Phys. \textbf{43}, 199 (1975) [erratum: Commun. Math. Phys.
\textbf{46}, 206 (1976)].


\bibitem{Adler:2001vs}
R.~J.~Adler, P.~Chen and D.~I.~Santiago, The generalized
uncertainty principle and black hole remnants, Gen. Rel. Grav.
\textbf{33}, 2101 (2001)
[arXiv:gr-qc/0106080 [gr-qc]].

\bibitem{Medved:2004yu}
A.~J.~M.~Medved and E.~C.~Vagenas, When conceptual worlds collide:
The GUP and the BH entropy, Phys. Rev. D \textbf{70}, 124021
(2004)
[arXiv:hep-th/0411022[hep-th]].

\bibitem{Nozari:2005ah}
K.~Nozari and S.~H.~Mehdipour, Gravitational uncertainty and black
hole remnants, Mod. Phys. Lett. A \textbf{20}, 2937 (2005)
[arXiv:0809.3144 [gr-qc]].


\bibitem{Chen:2014jwq}
P.~Chen, Y.~C.~Ong and D.~h.~Yeom, Black Hole Remnants and the
Information Loss Paradox, Phys. Rept. \textbf{603}, 1 (2015)
[arXiv:1412.8366 [gr-qc]].

\bibitem{Carr:2015nqa}
B.~J.~Carr, J.~Mureika and P.~Nicolini, Sub-Planckian black holes
and the Generalized Uncertainty Principle, JHEP \textbf{07}, 052
(2015)
[arXiv:1504.07637 [gr-qc]].

\bibitem{Bosso:2023aht}
P.~Bosso, G.~G.~Luciano, L.~Petruzziello and F.~Wagner, 30 years
in: Quo vadis generalized uncertainty principle?, Class. Quant.
Grav. \textbf{40}, 195014 (2023)
[arXiv:2305.16193 [gr-qc]].




\bibitem{Amelino-Camelia:2005zpp}
G.~Amelino-Camelia, M.~Arzano, Y.~Ling and G.~Mandanici,
Black-hole thermodynamics with modified dispersion relations and
generalized uncertainty principles, Class. Quant. Grav.
\textbf{23}, 2585-2606 (2006)
[arXiv:gr-qc/0506110 [gr-qc]].

\bibitem{Xiang:2006ei}
L.~Xiang, Dispersion relation, black hole thermodynamics and
generalization of uncertainty principle, Phys. Lett. B
\textbf{638}, 519 (2006).


\bibitem{Myung:2006qr}
Y.~S.~Myung, Y.~W.~Kim and Y.~J.~Park, Black hole thermodynamics
with generalized uncertainty principle, Phys. Lett. B
\textbf{645}, 393 (2007)
[arXiv:gr-qc/0609031 [gr-qc]].

\bibitem{Hai-Xia:2007}
H. Zhao, H. Li, S. Hu and R. Zhao, Generalized uncertainty
principle and black hole entropy of higher-dimensional de sitter
spacetime, Commun. Theor. Phys. \textbf{48}, 465 (2007).


\bibitem{Xiang:2009yq}
L.~Xiang and X.~Q.~Wen, Black hole thermodynamics with generalized
uncertainty principle, JHEP \textbf{10}, 046 (2009)
[arXiv:0901.0603 [gr-qc]].

\bibitem{Banerjee:2010sd}
R.~Banerjee and S.~Ghosh, Generalised uncertainty principle,
remnant mass and singularity problem in black hole thermodynamics,
Phys. Lett. B \textbf{688}, 224 (2010)
[arXiv:1002.2302 [gr-qc]].

\bibitem{Gangopadhyay:2013ofa}
S.~Gangopadhyay, A.~Dutta and A.~Saha, Generalized uncertainty
principle and black hole thermodynamics, Gen. Rel. Grav.
\textbf{46}, 1661 (2014)
[arXiv:1307.7045 [gr-qc]].

\bibitem{Gine:2020ves}
J.~Gin\'e, Modified Hawking effect from generalized uncertainty
principle, Commun. Theor. Phys. \textbf{73}, 015201 (2021).

\bibitem{Lutfuoglu:2021ofc}
B.~C.~L\"utf\"uo\u{g}lu, B.~Hamil and L.~Dahbi, Thermodynamics of
Schwarzschild black hole surrounded by quintessence with
generalized uncertainty principle, Eur. Phys. J. Plus
\textbf{136}, 976 (2021)
[arXiv:2110.01383 [gr-qc]].

\bibitem{Su:2022cwd}
H.~Su and C.~Y.~Long, Thermodynamics of the black holes under the
extended generalized uncertainty principle with linear terms,
Commun. Theor. Phys. \textbf{74}, 055401 (2022).


\bibitem{Yu:2024qzl}
B.~Yu and Z.~W.~Long, Black hole evaporation and its remnants with
the generalized uncertainty principle including a linear term,
Commun. Theor. Phys. \textbf{76}, 025404 (2024).








\bibitem{Liu:2001ra}
W.~B.~Liu and Z.~Zhao, The entropy calculated via brick-wall
method comes from a thin film near the event horizon, Int. J. Mod.
Phys. A \textbf{16}, 3793 (2001).

\bibitem{Li:2002xb}
X.~Li, Black hole entropy without brick walls, Phys. Lett. B
\textbf{540}, 9 (2002)
[arXiv:gr-qc/0204029 [gr-qc]].

\bibitem{Kim:2007nh}
W.~Kim, Y.~W.~Kim and Y.~J.~Park, Entropy of a charged black hole
in two dimensions without cutoff, Phys. Rev. D \textbf{75}, 127501
(2007)
[arXiv:gr-qc/0702018 [gr-qc]].

\bibitem{Nouicer:2007jg}
K.~Nouicer, Quantum-corrected black hole thermodynamics to all
orders in the Planck length, Phys. Lett. B \textbf{646}, 63 (2007)
[arXiv:0704.1261 [gr-qc]].

\bibitem{Kim:2007if}
Y.~W.~Kim and Y.~J.~Park, Entropy of the Schwarzschild black hole
to all orders in the Planck length, Phys. Lett. B \textbf{655},
172 (2007)
[arXiv:0707.2128 [gr-qc]].

\bibitem{Zhao:2009zzb}
R.~Zhao, Y.~Q.~Wu and L.~C.~Zhang, Entropy of a rotating and
charged black string to all orders in the Planck length, Chin.
Phys. B \textbf{18}, 1749 (2009).

\bibitem{Tang:2017wph}
H.~Tang, C.~Y.~Sun and R.~H.~Yue, Entropy of Schwarzschild-de
Sitter black hole with generalized uncertainty principle
revisited, Commun. Theor. Phys. \textbf{68}, 64 (2017).

\bibitem{Hong:2021xeg}
S.~T.~Hong, Y.~W.~Kim and Y.~J.~Park, GUP corrected entropy of the
Schwarzschild black hole in holographic massive gravity, Mod.
Phys. Lett. A \textbf{27}, 2250186 (2022)
[arXiv:2103.05755 [gr-qc]].



\bibitem{Scardigli:2014qka}
F.~Scardigli and R.~Casadio, Gravitational tests of the
generalized uncertainty principle, Eur. Phys. J. C \textbf{75},
425 (2015)
[arXiv:1407.0113 [hep-th]].


\bibitem{FaragAli:2015boi}
A.~Farag Ali, M.~M.~Khalil and E.~C.~Vagenas, Minimal Length in
quantum gravity and gravitational measurements, EPL \textbf{112},
20005 (2015)
[arXiv:1510.06365 [gr-qc]].

\bibitem{Vagenas:2017vsw}
E.~C.~Vagenas, S.~M.~Alsaleh and A.~Farag, GUP parameter and black
hole temperature, EPL \textbf{120}, 40001 (2017)
[arXiv:1801.03670 [hep-th]].

\bibitem{Contreras:2016xib}
E.~Contreras, F.~Villalba and P.~Bargue\~no, Effective geometries
and generalized uncertainty principle corrections to the
Bekenstein-Hawking entropy, EPL \textbf{114}, 50009 (2016)
[arXiv:1606.07281 [gr-qc]].



\bibitem{Anacleto:2020lel}
M.~A.~Anacleto, F.~A.~Brito, J.~A.~V.~Campos and E.~Passos,
Quantum-corrected scattering and absorption of a Schwarzschild
black hole with GUP, Phys. Lett. B \textbf{810}, 135830 (2020)
[arXiv:2003.13464 [gr-qc]].

\bibitem{Anacleto:2021qoe}
M.~A.~Anacleto, J.~A.~V.~Campos, F.~A.~Brito and E.~Passos,
Quasinormal modes and shadow of a Schwarzschild black hole with
GUP, Annals Phys. \textbf{434}, 168662 (2021)
[arXiv:2108.04998 [gr-qc]].

\bibitem{Jusufi:2022uhk}
K.~Jusufi and D.~Stojkovic, Theory and phenomenology of a
four-dimensional string\textendash{}corrected black hole, Universe
\textbf{8}, 194 (2022)
[arXiv:2203.10957 [gr-qc]].

\bibitem{Chemisana:2023fuk}
D.~Chemisana, J.~Gin\'e and J.~Madrid, Generalized Heisenberg
uncertainty principle due to the quantum gravitational effects in
the Schwarzschild spacetime, Nucl. Phys. B \textbf{991}, 116225
(2023).





\bibitem{Ong:2023jkp}
Y.~C.~Ong, A critique on some aspects of GUP effective metric,
Eur. Phys. J. C \textbf{83}, 209 (2023)
[arXiv:2303.10719 [gr-qc]].




\bibitem{MTW:1973}
C. W. Misner, K. S. Thorne and J. A. Wheeler, {\it Gravitation},
Pinceton University Press, Princeton and Oxford, 2017.

\bibitem{DInverno:1992}
R. D'Inverno, {\it Introducing Einstein's Relativity}, Clarendon
Press, Oxford, 1992.

\bibitem{Carroll:2004}
S. M. Carroll, {\it Spacetime and Geometry}, Addison Wesley, San
Franscico, 2004.

\bibitem{Hobson:2006}
M. P. Hobson, G. P. Efstathiou and A. N. Lasenby, {\it General
Realtivity: An Introduction for Physicists}, Cambridge University
Press, Cambridge, 2006.

\bibitem{Goswami:2019fyk}
R.~Goswami and G.~F.~R.~Ellis, Tidal forces are gravitational
waves, Class. Quant. Grav. \textbf{38}, 085023 (2021)
[arXiv:1912.00591 [gr-qc]].


\bibitem{Crispino:2016pnv}
L.~C.~B.~Crispino, A.~Higuchi, L.~A.~Oliveira and E.~S.~de
Oliveira, Tidal forces in Reissner\textendash{}Nordstr\"om
spacetimes, Eur. Phys. J. C \textbf{76}, 168 (2016)
[arXiv:1602.07232 [gr-qc]].

\bibitem{Shahzad:2017vwi}
M.~U.~Shahzad and A.~Jawad, Tidal Forces in Kiselev black hole,
Eur. Phys. J. C \textbf{77}, 372 (2017)
[arXiv:1706.00281 [gr-qc]].

\bibitem{Sharif:2018gaj}
M.~Sharif and S.~Sadiq, Tidal effects in some regular black holes,
J. Exp. Theor. Phys. \textbf{126}, 194 (2018).

\bibitem{Lima:2020wcb}
H.~C.~D.~Lima and L.~C.~B.~Crispino, Tidal forces in the charged
Hayward black hole spacetime, Int. J. Mod. Phys. D \textbf{29},
2041014 (2020)
[arXiv:2005.13029 [gr-qc]].

\bibitem{Hong:2020bdb}
S.~T.~Hong, Y.~W.~Kim and Y.~J.~Park, Tidal effects in
Schwarzschild black hole in holographic massive gravity, Phys.
Lett. B \textbf{811}, 135967 (2020)
[arXiv:2008.05715 [gr-qc]].


\bibitem{Li:2021izh}
J.~Li, S.~Chen and J.~Jing, Tidal effects in 4D
Einstein\textendash{}Gauss\textendash{}Bonnet black hole
spacetime, Eur. Phys. J. C \textbf{81}, 590 (2021)
[arXiv:2105.01267 [gr-qc]].

\bibitem{Vandeev:2021yan}
V.~P.~Vandeev and A.~N.~Semenova, Tidal forces in Kottler
spacetimes, Eur. Phys. J. C \textbf{81}, 610 (2021)
[arXiv:2204.13203 [gr-qc]].

\bibitem{Vandeev:2022gbi}
V.~P.~Vandeev and A.~N.~Semenova, Deviation of non-radial
geodesics in a static spherically symmetric spacetime, Eur. Phys.
J. Plus \textbf{137}, 185 (2022)
[arXiv:2204.13200 [gr-qc]].

\bibitem{Madan:2022spd}
S.~Madan and P.~Bambhaniya, Tidal force effects and periodic
orbits in null naked singularity spacetime, [arXiv:2201.13163
[gr-qc]].

\bibitem{Liu:2022lrg}
J.~Liu, S.~Chen and J.~Jing, Tidal effects of a dark matter halo
around a galactic black hole, Chin. Phys. C \textbf{46}, 105104
(2022)
[arXiv:2203.14039 [gr-qc]].

\bibitem{LimaJunior:2022gko}
H.~C.~D.~Lima, Junior, M.~M.~Corr\^ea, C.~F.~B.~Macedo and
L.~C.~B.~Crispino, Tidal forces in dirty black hole spacetimes,
Eur. Phys. J. C \textbf{82}, 479 (2022)
[arXiv:2205.13569 [gr-qc]].


\bibitem{Abbas:2023gap}
G.~Abbas and M.~Asgher, Tidal effect in charged black hole
enclosed by thin accretion disc in Rastall gravity, New Astron.
\textbf{99}, 101967 (2023).


\bibitem{Toshmatov:2023anz}
B.~Toshmatov and B.~Ahmedov, Tidal forces in parametrized
spacetime: Rezzolla-Zhidenko parametrization, Phys. Rev. D
\textbf{108}, 084035 (2023).



\bibitem{Rashidi:2015rro}
R.~Rashidi, Generalized uncertainty principle and the maximum mass
of ideal white dwarfs, Annals Phys. \textbf{374}, 434 (2016)
[arXiv:1512.06356 [gr-qc]].

\bibitem{Ong:2018zqn}
Y.~C.~Ong, Generalized Uncertainty Principle, Black Holes, and
White Dwarfs: A Tale of Two Infinities, JCAP \textbf{09}, 015
(2018)
[arXiv:1804.05176 [gr-qc]].

\bibitem{Mathew:2020wnx}
A.~Mathew and M.~K.~Nandy, Existence of Chandrasekhar's limit in
generlized uncertainty white dwarfs, R. Soc. Open Sci. \textbf{8},
210301 (2021)
[arXiv:2002.08360 [gr-qc]].




\bibitem{Dvali:2010bf}
G.~Dvali and C.~Gomez, Self-Completeness of Einstein Gravity,
[arXiv:1005.3497 [hep-th]].

\bibitem{Isi:2013cxa}
M.~Isi, J.~Mureika and P.~Nicolini, Self-Completeness and the
Generalized Uncertainty Principle, JHEP \textbf{11}, 139 (2013)
[arXiv:1310.8153 [hep-th]].

\bibitem{Li:2016yfd}
X.~Li, Y.~Ling, Y.~G.~Shen, C.~Z.~Liu, H.~S.~He and L.~F.~Xu,
Generalized uncertainty principles, effective Newton constant and
the regular black hole, Annals Phys. \textbf{396}, 334 (2018)
[arXiv:1611.09016 [gr-qc]].


\bibitem{Alencar:2023wyf}
G.~Alencar, M.~Estrada, C.~R.~Muniz and G.~J.~Olmo, Dymnikova
GUP-corrected black holes, JCAP \textbf{11}, 100 (2023)
[arXiv:2309.03920 [gr-qc]].



\end{thebibliography}
\end{document}